# Street Marketing:
# How Proximity and Context drive Coupon Redemption

*Sarah Spiekermann, Matthias Rothensee and Michael Klafft*

**Abstract**
**Purpose -** In 2009, U.S. coupons set a new record of 367 billion coupons distributed. Yet, while coupon distribution is on the rise, redemption rates remain below 1%. This article shows how recognizing context variables, such as proximity, weather, part of town and financial incentives interplay to determine a coupon campaigns' success.
**Design/methodology/approach –** The paper reports an empirical study conducted in co-operation with a restaurant chain: 9.880 Subway coupons were distributed under different experimental context conditions. Redemption behavior was analyzed with the help of logistic regressions.
**Findings –** We found that even though proximity drives coupon redemption, city center campaigns seem to be much more sensitive to distance than suburban areas. The further away the distribution place from the restaurant the less does the amount of monetary incentive determine the motivation to redeem.
**Practical implications –** When designing a coupon campaign for a company, coupon distribution should not follow a 'one-is-good-for-all-strategy' even for one marketer within one product category. Instead each coupon strategy should carefully consider contextual influence.
**Originality –** This article is the first to our knowledge that systematically investigates the impact of context variables on coupon redemption. We focus on context variables that electronic marketing channels will be able to easily incorporate into personalized mobile marketing campaigns.

**Keywords** Coupon campaign, Mobile marketing, Coupon redemption

**Paper type** Research paper

1. Introduction

In 2009, U.S. coupons set a new record of 367 billion coupons distributed (CouponInfoNow.com, 2010). Yet, while coupon distribution is on the rise, redemption rates remain below 1% (CouponInfoNow.com, 2010).

However, the coupon industry is changing: Consumers are overwhelmed by the choice of coupons they receive. In-store coupons are becoming more popular. Online models are rapidly embraced. And, most important, mobile phone couponing finally seems to be taking off: First mobile Bluetooth campaigns have been launched successfully (Leek;Christodoulides, 2009) and research is intensifying around "virtual purses" that could collect coupons along the way (Ferscha;Swoboda;Wimberger, 2009). In 2006, spending on mobile ads already amounted to $ 871 million worldwide and this sum is expected to grow to $ 11,4 billion by 2011 (The Economist, 2007). Against this background, a thorough understanding of critical success factors for online and offline couponing campaigns becomes increasingly important.

Since the 1960s, marketing researchers and the industry have continued to build up knowledge on how to stir couponing campaigns (Ferris, 2007; Lichtenstein;Netemeyer;Burton, 1990; Reibenstein;Traver, 1982; Ward;Davis, 1978). By now it is common knowledge that the face value of a coupon or the discount promised drive its redemption (Chakraborty;Cole, 1991; Reibenstein;Traver, 1982). Equally, distribution media and the size of coupon drops are important for coupon redemption (Reibenstein;Traver, 1982), individual coupon proneness (Lichtenstein et al., 1990) or even retail personnel perception of coupon use (Brumbaugh;Rosa, 2009). Ideally, scholars would like to predict the success of couponing campaigns based on a few aggregate data points on a campaign (Musalem;Bradlow;Raju, 2008). Yet, with the trend moving towards point-of-sale campaigns as well as mobile phone couponing, new distribution dynamics and opportunities are on the rise. One key opportunity to foster couponing success is to personalize campaigns on the basis of *context* variables (Dickinger;Kleijnen, 2009).

Context can be understood as the "circumstances, background or setting" in which an action takes place (Bradley;Dunlop, 2005). While context modeling is an area much researched in computer science today (Agre, 2001; Coutaz;Crowley;Dobson et al., 2005), few studies exist on the economic implications of context integration into personalized advertisements. For instance: Is it always best to distribute coupons to those customers closest by? And, is proximity a driver of coupon redemption in all locations alike? How does the environment such as population, weather or the degree of urbanization impact redemption behavior? What interplay is there between discount rate and neighborhood? Has ad timing the same impact in a cozy suburban main-street than in a busy tourist area?

In this article we present a highly controlled experiment conducted in co-operation with the restaurant chain Subway. Subway is part of the restaurant industry that is currently seeing a turnover of $ 566 billion (Association, 2009), over $ 110 billion of which are spent in comparable fast food chains (Schlosser, 2001). We observed redemption behavior for 9.880 distributed coupons for which the proximity, face value, town area and distribution time were systematically varied to observe their impact. The insights gained for the product category go beyond the expected: We find that one context variable alone, such as proximity, does not drive coupon redemption. Instead context variables seem to interplay. For example, city center campaigns seem to be more sensitive to the proximity of the redemption place than suburban areas.

## 2. Theoretical background and hypotheses

### 2.1. Some hypotheses on the power of physical proximity

When coupons or ads are delivered to customers through physical or electronic channels an intuitive judgment is that the closer the point of delivery to the point of redemption, the higher will be the campaign's success.

Not surprisingly, scientists and mobile marketing experts see location specificity as one of the most important variables for successful mobile marketing (Leppäniemi;Karjaluoto, 2005; Mobile Marketing Association, 2007). 85% of mobile marketing professionals consider the recognition of location in mobile ads as very important or important. 45% even view it as critical for the success of a campaign (Marketing Week, 2001).

Transaction utility theory in economics supports this intuition. It suggests that increasing the proximity of coupon delivery to a place of redemption directly drives customer benefit, because the price advantage offered to a prospect customer is not undermined by any transaction cost to travel to the place of redemption (Thaler, 1983). Marketing theorists have employed transaction utility theory in former couponing studies to show how 'value consciousness' of consumers drives redemption behavior independent of their 'coupon proneness' (Lichtenstein et al., 1990).

Against this background, we hypothesize:

**H1:** The bigger the proximity of coupon delivery to the place of redemption, the higher the coupon redemption rate.

A negative side effect implicit in hypothesis 1 is that high proximity campaigns could lead to windfall gains for consumers. In economic theory unexpected gains are referred to as 'windfall gains'. In the context of couponing campaigns, consumers would benefit from windfall gains if they had bought a respective product even without owning a coupon for it. The face value of the discount unexpectedly received is the windfall gain.

Windfall gains have been identified as a particular challenge for in-store couponing promotion (Ha;Hyun;Pae, 2006). Buyers who planned to purchase the discounted product at the given moment and at the regular price anyway benefit from a discount while the seller does not gain a new customer or prospect.

The question underlying hypothesis 2 is therefore whether proximity of coupon delivery systematically reinforces a product choice already made, or triggers by-passers' active decision for a product advertised. It seems logical to argue that if a coupon is distributed closer to the place of redemption, it will more likely be given out to people who are on their way to the store anyways and would have bought the advertised product even without a coupon. Therefore we hypothesize:

**H2:** The bigger the proximity of coupon delivery to the place of redemption, the higher are customer windfall gains.

Hypothesis 1 and 2 analyze the value of proximity in relation to transaction costs for prospect consumers. Economic theory on retail economics suggests, however, that the real power of advertisements resides in the fact that consumers outside of a retailer's visible location radius can receive attractive price information. This price information could draw competitors' clientele to a retailer's premises (Balasubramanian;Peterson;Jarvenpaa, 2002). A supposition of this argument is that the transaction cost of traveling to the place of redemption is justified by the financial gain signaled to consumers.

The importance of financial gain promised to prospect consumers has also been proven by marketing research on coupon redemption. Several studies have shown that the face value of a coupon or the discount promised drive its redemption (Chakraborty;Cole, 1991; Reibenstein;Traver, 1982). Against this background we expect a relationship between coupon face value and redemption ratio:

**H3:** The higher the coupon's face value, the higher the redemption rate.

Yet, it has not been investigated whether coupon face value (transaction benefit) and proximity (transaction cost) have a systematic relationship. Thus, can distance from the store be compensated by coupon face value? And, is the relationship between the face value of a coupon and distance to the place of redemption a linear one? We state the following hypothesis:

**H3a:** The impacts of proximity of coupon delivery and face value of a coupon on its redemption rate are compensatory.

## 2.2. Proximity and Coupon Safekeeping

When investigating drivers of coupon redemption scholars found that expiration dates have a significant effect on redemption (Inman;McAlister, 1994). Due to the fear of regretting non-redemption, most coupons are turned in early on in a campaign period. Only towards the end of a campaign's duration a renewed smaller redemption peak can be observed (Inman;McAlister, 1994). This finding has been important for the prediction of promotion stocks and campaign expenses.

In the context of mobile marketing a related question arises, which is to ask how the proximity of a coupon's delivery to the place of redemption impacts peoples' propensity to redeem immediately or rather keep the coupon for redemption at a later point in time. Incurring the technical cost of mobile localization only makes sense in a mobile campaign if the location triggers immediate redemption. Otherwise, traditional media could also be used for the advertisement's distribution or at least the cost of localization could be saved. Seen the context specificity and relevance aspired through precise localization, we expect that proximity discourages safekeeping of coupons. Studies building on transaction-utility would support this notion (Lichtenstein et al., 1990). Equally, the theory of immediate gratification would suggest that the immediate availability of an offer very close by would tempt a person to redeem even if he or she had not previously intended to redeem at further distances (O'Donoghue;Rabin, 2000). We therefore hypothesize:

**H4:** The higher the proximity of coupon delivery the less likely becomes coupon safekeeping for later.

## 2.3. Coupon redemption in urban vs. suburban areas

A major assumption of the hypotheses made so far is that the impact of proximity, face value and safekeeping on redemption behavior are the same for all places at which mobile advertisements occur. Yet, some early studies on couponing success suggest that the area of country can influence redemption behavior (Reibenstein;Traver, 1982). People who potentially receive coupons are different in each city area (e.g., tourists in centres vs. local population in suburban areas), they travel to these areas for different purposes (e.g., work vs. leisure), and differ in their flexibility regarding a change of plans, such as where to go for dinner. Yet, rarely has any research tested the effect of a part of town on coupon redemption. As a result, we expect all hypothesized relationships to hold true no matter where a mobile marketing campaign is conducted. This leads to five additional hypotheses:

**H5-1:** The higher the proximity of coupon delivery is to the place of redemption, the higher the coupon redemption rate in both city centre and suburban area.

**H5-2:** The higher the proximity of coupon delivery to the place of redemption, the higher the customer windfall gains in both city centre and suburban area.

**H5-3:** The higher the coupon's face value, the higher the redemption rate in both city centre and suburban area.

**H5-3a:** The impacts of proximity of coupon delivery and face value of a coupon on it's redemption rate are compensatory in both city centre and suburban area.

**H5-4:** The higher the proximity of coupon delivery the less likely becomes coupon safekeeping in both city centre and suburban area.

## 3. Methodology

To test the impact of proximity we conducted an experiment in co-operation with the fast food chain Subway. We organised a manual couponing campaign in the vicinity of two franchise restaurants in a major European city. One of these restaurants was located in the city centre and close to the main tourist attractions of Berlin. The second one was situated in the main shopping street of one of Berlin's suburbs.

9880 coupons with different monetary incentives were distributed in varying distances from the two restaurants. The distances between the places of distribution from the place of redemption were 10 meters, 250 meters, 400 meters or 800 meters. The distances were chosen to be roughly similar to what can be expected from different localization technologies supporting mobile advertisement. A-GPS or Bluetooth, for example, can localize customers directly in front of stores (10 metres and less). Wireless LAN and dense GSM network infrastructures identify customers reliably within a 250 m precision radius. And 400 to 800 metres localization precision can be obtained from most mobile networks even if there is no high cell density. At the distance of 10 and 250 meters the restaurant was in peoples' range of vision.

The savings offered were either €1, €1,50 or €2 on any meal chosen by the customer. The coupons were valid for one month. Nominal discounts were used instead of relative ones in order to make the benefit of the coupon as transparent as possible (Chen;Kent;MonroeYung-Chien, 1998).

We chose a manual campaign over a handset based campaign, because GPS enabled phones or bluecasting kiosks necessary for mobile delivery are not as widely available yet to ensure equal access to coupons. Using only a selected group of participants owning high-end phones would have potentially distorted the sample's observed behaviour. Moreover, mobile phone interfaces can vary considerably in their usability and could have led to different perceptions and transaction costs associated with the electronic coupon.

The point could be made that electronic coupons delivered to personal mobile devices are delivered neutrally through the technology. Or in other words, there is no experimenter bias, like an ad distributor's looks or level of empathy that may influence redemption behaviour. To minimize such distributor effects on redemption, all distributors wore the same clothes with Subway's corporate layout. They were specially trained prior to distribution to offer the coupons to all passers in a non-intrusive manner and to not engage in any potentially convincing behaviours or conversations with potential customers.

All coupons were numbered to make them traceable, and time and place of distribution were recorded during the entire experiment. Additional context factors were controlled through the link of the coupon ID with external information. These included information on the weather situation, information on the precise time of redemption, and customers' total consumption. Furthermore, we distributed a short questionnaire in the two restaurants in order to control for some factors potentially influencing the observed behavior - for example, patronage.

Of the 171 customers who redeemed a coupon, 116 customers answered the questionnaire upon coupon redemption. This represents a participation rate of 63%. Table 1 summarizes the details on the distribution plan.

Table 1: Number of coupons distributed per discount category and distribution distance

|  | **City Centre** | | | | **Suburb** | | | |
|---|---|---|---|---|---|---|---|---|
| Discount | 1 € | 1,50 € | 2,00 € | Total | 1 € | 1,50 € | 2,00 € | Total |
| Distance | | | | | | | | |
| 10 m | 200 | 169 | 293 | 662 | 285 | 264 | 262 | 811 |
| 250 m | 301 | 176 | 247 | 724 | 316 | 316 | 306 | 938 |
| 400 m | 257 | 373 | 612 | 1242 | 441 | 436 | 473 | 1350 |
| 800 m | 574 | 921 | 643 | 2138 | 518 | 542 | 512 | 1572 |

## 4. Results

In order to investigate the determinants of coupon redemption we report descriptive statistics and logistic regression analyses (Menard, 1995) of the actual coupons redeemed. In addition, answers from the post-questionnaire are analysed. The results section covers analyses on the effects of proximity (H1, H2) and face value (H3) of the coupons, as well as their joint influence on redemption (H3a), windfall gains (H2), and promptness of redemption (H4). Finally, the analyses are repeated with a view on the distinction between city centre and suburban area (H5).

### 4.1. Overall findings on the power of proximity

Table 3 shows that the higher the distance to the restaurant is the lower redemption rates fall. The redemption rates in 10m distance were 5 times as high as in 800m distance. In order to statistically prove the significance of this result, and to allow for joint analyses of other factors along with the distance to the restaurant, a logistic regression analysis was computed. Logistic regression predicts whether a coupon will be redeemed or not based on several predictor variables. For the entire sample a logistic regression analysis was computed that took into account distance from the restaurant, face value of the voucher, and time of day. Furthermore, the location of distribution was included as a binary dummy variable (city centre vs. suburb). The regression equation was computed in a backward stepwise manner. B-coefficients can be interpreted in terms of direction and magnitude of change in the logit of the outcome as a result of a one-unit change of the predictor variable. The column 'Wald' indicates whether the estimate is significantly different from zero and therefore has to be regarded as important in the regression. In order to show that the inclusion of additional parameters improves the prediction of customer behavior, the actual regression equation has to prove more effective than the baseline model (which, in this case, is the outcome which appeared most often, i.e. "non-return" of the vouchers). The regression equations improved the fit of the baseline model ($c^2_{baseline}$ = 1710.55; $c^2_{model}$ = 1584.11; $p < .01$). Thus, the variables included in the regression equation make a meaningful contribution to explaining coupon redemption. Results of this analysis are displayed in table 3.

Table 2: Coupon redemption, windfall gains, and promptness of redemption by distribution distance

|       | Coupons... | | | Windfall Gains | | Promptness of Redemption | |
|-------|------------|----------|------|----------------|---------------|-------------|-----------|
|       | distributed | redeemed | %    | % windfall gains | % plan change | % Immediate | % Delayed |
| 10 m  | 1473       | 68       | 4.62 | 35.3           | 64.7          | 51.5        | 48.5      |
| 250 m | 1664       | 36       | 2.16 | 36.1           | 63.9          | 52.8        | 47.2      |
| 400 m | 2592       | 35       | 1.35 | 42.9           | 57.1          | 28.6        | 71.4      |
| 800 m | 3710       | 32       | 0.86 | 34.4           | 65.6          | 21.9        | 78.1      |

The regression equation shows that in accordance with hypothesis 1 there is a strong positive relationship between proximity of coupon delivery and redemption. This result mirrors what was found in the redemption rates reported in table 3. Thus, hypothesis 1 is accepted on the basis of the present analyses.

Table 3: Overall Logistic Regression on Coupon Redemption

|            | B      | S.E.  | Wald    | df | Sig.  |
|------------|--------|-------|---------|----|-------|
| Distance[a] | -0.215 | 0.03  | 51.399  | 1  | 0.000 |
| Discount[b] | 1.192  | 0.214 | 31.145  | 1  | 0.000 |
| Day Time[c] | -0.062 | 0.028 | 4.988   | 1  | 0.026 |
| Town Area  | 0.444  | 0.164 | 7.341   | 1  | 0.007 |
| Constant   | -5.493 | 0.481 | 130.279 | 1  | 0     |

(*Note:* [a] in 100 m, [b] in Euro, [c] in hours,)

**Windfall gains.** The finding on hypothesis 1 suggests that it makes most sense for marketers to distribute coupons directly in front of stores in city centers to boost sales. This simple advice, however, could be contradicted by windfall gains for customers. That is, if customers were to visit the restaurant anyway, the restaurant company would win nothing if the goal of the campaign was attracting new customers that had not planned to eat at the issuing restaurant. To investigate this issue, redeemers were asked to state whether they visited the restaurant because of the received coupon or not. The percentages of windfall gains are provided in table 2. As can be seen, there is no systematic relationship between distribution distance and percentage of customers visiting the restaurant due to receipt of the coupon. In other words, irrespective of distribution location, the marketer can expect to win roughly two thirds of redeemers because of the coupon that had not planned to visit his premises. The remaining third of participants will profit from a windfall gain. Hypothesis 2 is therefore rejected: the proximity of coupon redemption is unrelated to the percentage of windfall gains.

**Face value: monetary and distance thresholds.** Transaction costs occurring due to a walking distance may only be one side of the medal when explaining coupon redemption. The other side may regard the monetary incentive. Figure 1 shows that coupons with a higher face value are more readily redeemed, thereby confirming hypothesis 3. Interestingly, the € 1 or € 1.50 incentives behave similarly and cluster on one side versus the € 2 incentive on the other. Providing prospect customers with a € 2 incentive amounted to an average 35% discount on the total average meal price of € 5.75 and led to higher redemption than the lower face values. In the case of coupons for Subway meals the monetary threshold of coupon effectiveness lies between 1.5 and 2 EUR. This threshold will most likely depend on the average meal price at the respective restaurant. However, this cannot be proven based on the setup of the present study.

Interestingly, looking at the interplay between monetary incentive and proximity also allows for the identification of distance thresholds for the coupons. A visual inspection of figure 1 makes clear that beyond 400 m distance from the restaurant – thus beyond viewing distance of the restaurant - the face value of the coupon hardly has any impact on the redemption rate. For both 1 EUR and 1.5 EUR vouchers the point of steepest decrease in effectiveness lies between 10 and 250 m. But for the more valuable coupons this tipping point shifts – the likelihood of redemption is significantly lower at 400 m and above compared with distances of 250 m and below. Thus, a higher monetary incentive serves to push back the threshold of low coupon effectiveness. Generally, it is visible that the further away the distribution place the less does the amount of monetary incentive gain influence the motivation to redeem.

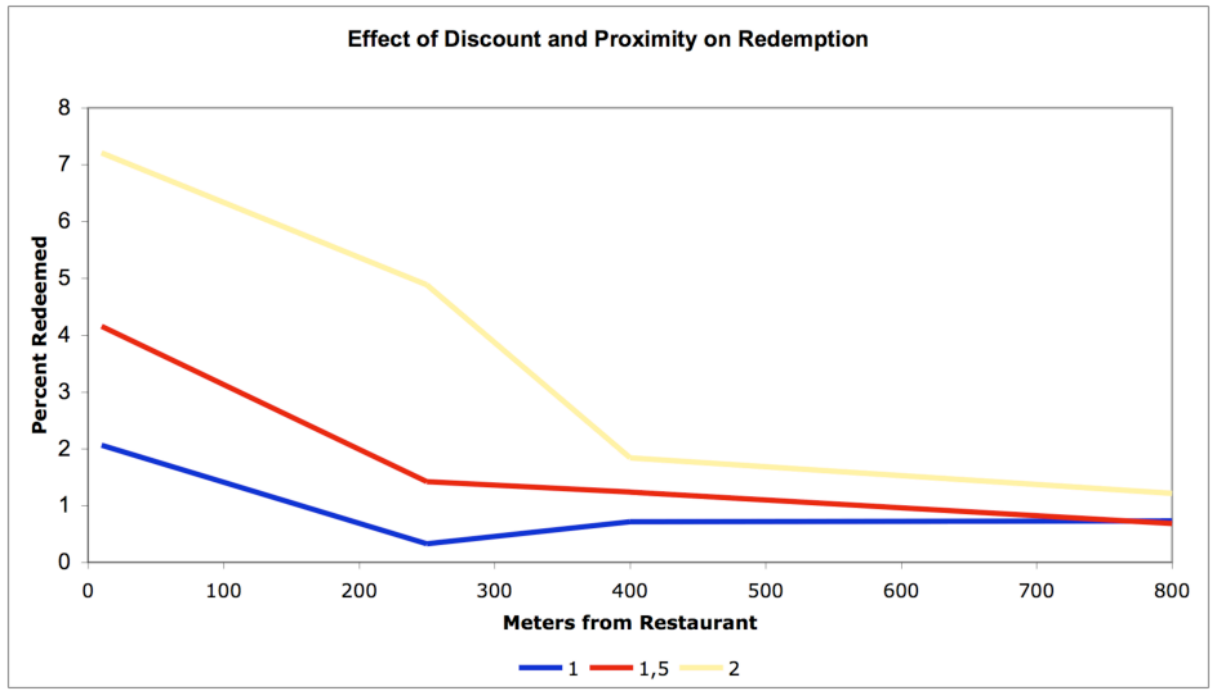

Figure 1: Effect of Discount and Proximity on Redemption

An additional logistic regression analysis was computed in order to find out whether the aforementioned compensatory effect of a monetary amount on effective distance of the coupons' distribution from the shop can be statistically proven. The previously described logistic regression analysis was re-calculated including the product of distance and face value as an additional predictor of redemption. The analysis reveals that an equally well-fitting model ($c^2_{model}$ = 1581.91) is established when the product of distance and face value are included. The unmultiplied proximity is excluded by the computation algorithm in favor of the product of distance and discount value. This means that, weighted by the face value, the distance is a better predictor of redemption than unweighted. On the basis of this finding the exploratory Hypothesis 3a is confirmed. Moreover, as shown in table 4, the face value of the coupon has a significant effect on coupon redemption over and above the product with distance, whereas distance alone does not.

Table 4: Alternative Overall Logistic Regression on Coupon Redemption

|  | B | S.E. | Wald | df | Sig. |
|---|---|---|---|---|---|
| Discount | 1.568 | 0.217 | 52.382 | 1 | 0 |
| Day Time | -0.06 | 0.028 | 4.68 | 1 | 0.031 |
| Town Area | 0.438 | 0.164 | 7.113 | 1 | 0.008 |
| *Discount*Distance* | -0.127 | 0.018 | 52.508 | 1 | 0 |
| Constant | -6.134 | 0.49 | 156.815 | 1 | 0 |

**Immediate redemption.** When it comes to the relationship between proximity and promptness of redemption we expected that a small distance would increase the probability of people immediately redeeming their coupons. As the rightmost columns of table 2 show, this expectation was correct. Over 50% of the coupon recipients redeemed it right after reception at 10 meters distance from the store. The same is true for the 250 m distance. At 400 and 800 m, the rate of customers with immediate redemption decreases down to 20-30%. Consequently, hypothesis 4 is confirmed by the present data.

### 4.2. Coupon redemption in urban vs. suburban neighbourhoods

As a final step in the analyses, all preceding analyses have been repeated with a view to differences between the city centre and the suburban areas. As can be read from the overall regression equation in table 3, the influence of the type of urbanization on redemption of coupons is significant. Therefore, separate logistic regression equations were computed for the city center and suburban area participants, respectively. The results are provided in table 5. The separated logistic regressions were able to significantly improve the fit of the baseline model, both in the centre of the city ($c^2_{baseline}= 1008.88$; $c^2_{model} = 875.99$; $p < .01$) and in the suburban area ($c^2_{baseline}= 693.31$; $c^2_{model} = 684.42$; $p < .01$). It is noteworthy that the increase in the goodness of fit of the model was greater in the city centre. This can be explained by the slightly higher variation in voucher returns in this area, and therefore a slightly larger amount of variance can be explained. The comparatively high Chi-Square values are an artefact of the sample size.

Table 5 – Coupon Redemption Determinants in the City Centre vs. Suburban Setting

|  | City Centre | | | | | | Suburban Area | | | | |
|---|---|---|---|---|---|---|---|---|---|---|---|
|  | B | S.E. | Wald | df | Sig. |  | B | S.E. | Wald | df | Sig. |
| Distance | -0.281 | 0.04 | 49.358 | 1 | 0 | Distance | -0.103 | 0.045 | 5.33 | 1 | 0.021 |
| Discount | 1.636 | 0.311 | 27.735 | 1 | 0 | Discount | 0.571 | 0.309 | 3.409 | 1 | 0.065 |
| Day Time | -0.109 | 0.041 | 7.211 | 1 | 0.007 | Day Time | -- | -- | -- | -- | -- |
| Constant | -5.024 | 0.604 | 69.179 | 1 | 0 | Constant | -4.72 | 0.53 | 79.446 | 1 | 0 |

**Proximity and neighbourhood** The influence of proximity of distribution on the redemption rate is significant at the 1% level for the city centre. For suburbs the distribution proximity is slightly less important and approaches significance at the 5% level. Thus, city center campaigns seem to be more sensitive to coupon distribution proximity than suburban areas. As figure 2 illustrates, highest proximity to a place of redemption (10 meters) can yield redemption rates that are more than three times as high in city centers than in suburban areas. The further away coupon distribution is from the place of redemption the higher the convergence of success rates for suburbs and city centers. This result is graphically depicted in figure 2. As proximity is important in both areas of town, hypothesis 5-1 is confirmed on the basis of the present data.

Table 6 – Coupon redemption, windfall gains, and promptness of redemption by distribution distance and type of urbanization

| City Centre | Coupons... distributed | redeemed | % | Windfall Gains % windfall gains | % plan change | Promptness of Redemption % immediate | % delayed |
|---|---|---|---|---|---|---|---|
| 10 m | 662 | 49 | 7.4 | 30.6 | 69.4 | 53.1 | 46.9 |
| 250 m | 724 | 23 | 3.2 | 21.7 | 78.3 | 52.2 | 47.8 |
| 400 m | 1242 | 17 | 1.4 | 17.6 | 82.4 | 52.9 | 47.1 |
| 800 m | 2138 | 16 | 0.7 | 18.8 | 81.3 | 37.5 | 62.5 |

| Suburban Area | Coupons... distributed | redeemed | % | Windfall Gains % windfall gains | % plan change | Promptness of Redemption % immediate | % delayed |
|---|---|---|---|---|---|---|---|
| 10 m | 811 | 19 | 2.3 | 47.4 | 52.6 | 47.4 | 52.6 |
| 250 m | 940 | 13 | 1.4 | 61.5 | 38.5 | 53.8 | 46.2 |
| 400 m | 1350 | 18 | 1.3 | 66.7 | 33.3 | 5.6 | 94.4 |
| 800 m | 1572 | 16 | 1.0 | 50.0 | 50.0 | 6.3 | 93.7 |

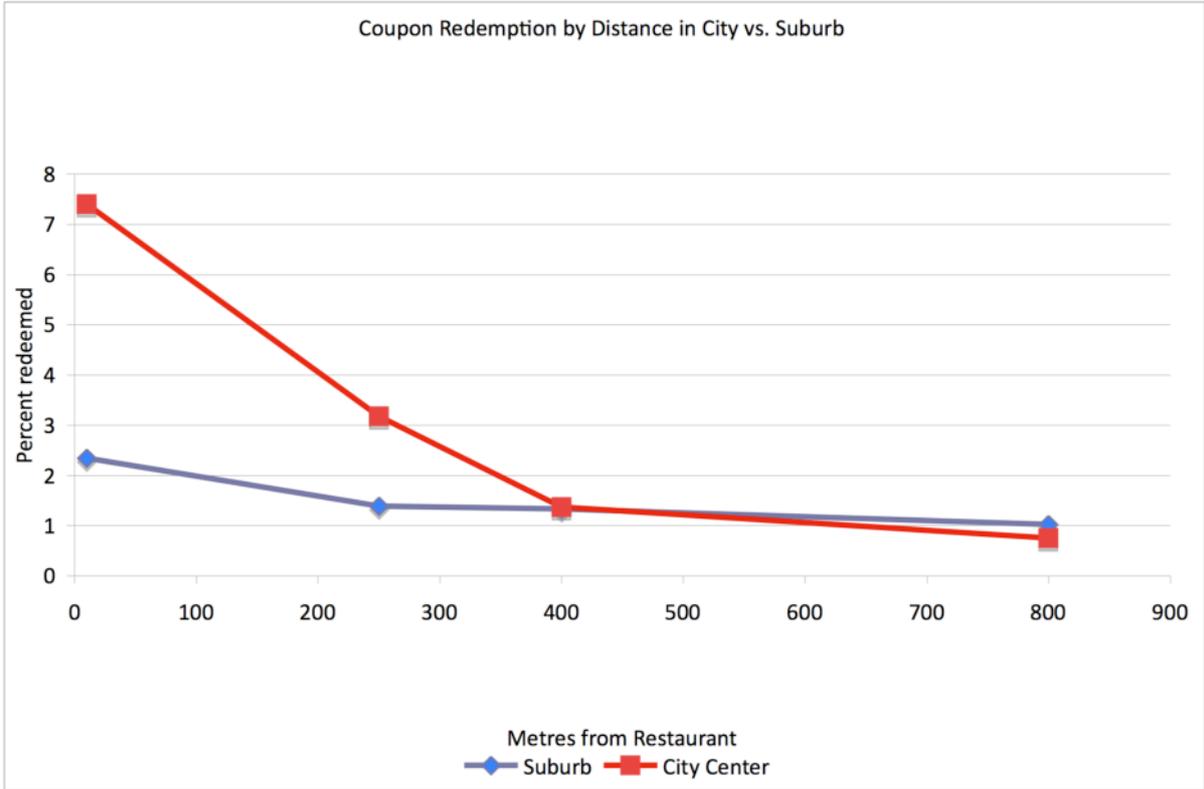

Figure 2: Effect of Proximity on Redemption in City Centre vs. Suburban Area

**Windfall gains and neighbourhood.** When comparing suburbs with city centers a strong difference in terms of windfall gains can be observed. The respective percentages are given in table 6. On the suburban main street the population of redeemers consists to a higher degree of those enjoying a windfall gain, i.e., with the intention to eat at Subway anyways. The percentage of windfall gains is up to 3 times higher in the suburban setting than in the city centre. Consequently, Hypothesis 5-2 is rejected. One reason for this observation could be that restaurant competition in a city centre is much higher than in a suburb. Thus people may be less planning and more spontaneous in their restaurant choice. In the suburban context consumers are probably more determined in where to have lunch or dinner. Therefore, the ones who redeem the coupons are only those who had planned to eat at Subway's beforehand.

**Monetary incentive and neighbourhood.** The effect of the monetary incentive on the redemption rate can universally be found in urban and suburban settings alike, thus supporting H5-3. Nevertheless, as can be seen in figure 3, it is less pronounced in the suburban setting. Consequently, it only approaches statistical significance in the separated logistic regression analysis reported in table 5. The compensatory effect of face value on effective distance is only pronounced at the city center. Redemption rate in the suburban area declines linearly. These visual inspection insights cannot be confirmed in the logistic regression due to sample size constraints (too few redemptions in the suburbs). Thus, hypothesis 5-3a is rejected on the basis of this finding.

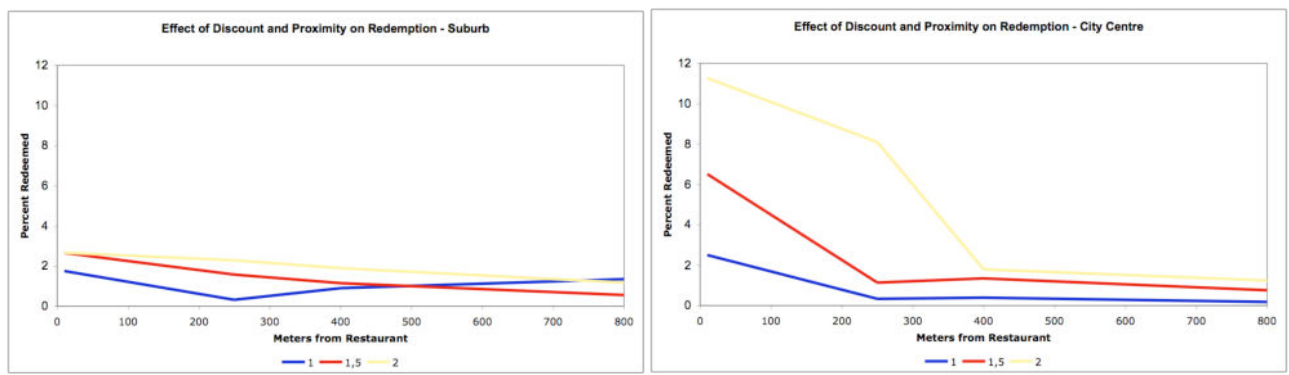

Figure 3: Effect of Discount and Proximity on Redemption in City Centre vs. Suburban Area

**Immediate redemption and neighbourhood.** Redemption rates separated by area of distribution are provided in the rightmost columns of table 6. As can be seen, in sight distance of the restaurant no differences between city centre and suburban area exist regarding promptness of redemption. Approximately half of the redeeming customers do so immediately after coupon receipt, the other half keeps the coupon for later redemption. At higher distances from the restaurant, however, the rate of immediate redemption steeply drops in the suburban area, but not in the city centre. Presumably, this occurs because in the city centre more coupons are distributed to passersby on a more leisurely stroll through the vicinity than in the suburban area. Thus, hypothesis 5-4 is rejected in the present study.

## 5. Managerial Implications

A main result from the experiment conducted is that the proximity of coupon delivery to a place of redemption is a significant driver of a couponing campaign's success. Therefore, the more precise the localization technology used as part of a campaign, the better it is. Yet, in city centers this dynamic is much more obvious than in suburban areas.

High proximity campaigns can create an area of conflict with high windfall gains for consumers. Our data suggests that for the present setup – a 20-30% discount on a 5-6 EUR meal at a fast food restaurant in Berlin – there is no higher degree of windfall gains at closer proximity than with consumers further off the premises. Nevertheless, as this situation might differ in different contexts, deciding on the degree of proximity should depend on the goal of a mobile marketing campaign. If a campaign is organized in order to move overstocked items to avoid shrinkage, to respond to competitive pressure, or to launch a new product for which word of mouth needs to be spread, then high proximity campaigns in suburban areas promise a high traffic to marketers' premises. If, however, the campaigns' goal is rather to win new customers, for trial or to encourage an unplanned (re)purchase, then campaigns in city centre areas seem more fruitful, because they reach more consumers who would not have come otherwise.

Another observation made in our experiment is that proximity can compensate for financial gains offered to customers. In our case, a low incentive of below 20% discount provided to prospect consumers directly in front of the restaurant more than doubled redemption rates. The most powerful effect of proximity can be achieved when combined with generous financial incentives. In our experiment, doubling the financial incentive at 10 meters from €1 to €2 led to a more than threefold increase of the redemption rate. Especially when launching a new retail outlet, or completely new brand, or product, such mobile campaigns could therefore be considered.

An important question is whether our results are transferable to other product domains than gastronomy. Certainly, price sensitivity and thus discount levels offered are highly specific to a product domain even though the overall relationship between proximity, price and redemption is probably similar (Chen et al., 1998). An interesting research question would be whether there is a phenomenon describable as 'product domain proximity sensitivity' that would reflect peoples' perception of acceptable distance to a point of sale for different price and product categories. Anecdotal evidence shows that people are willing to accept very high distances to a seller when such high involvement and high priced goods as automobiles are concerned. Perhaps there are even product categories where proximity is not an issue at all; such as when people travel the world for rare pieces of art.

Last but not least we identified significant differences in redemption behavior in suburbs versus city centres. Prospects in suburbs are less sensitive to proximity, slightly less sensitive to the level of financial incentive, change their eating plans less readily upon coupon reception and are also more likely to keep their coupons for later redemption. One explanation for these results is that the suburban areas find typically less fast food restaurants, therefore the lower competition of the Subway restaurant in focus might have triggered a more focused coupon redemption. On the other hand it is also possible that people receive coupons in between different activities. A more leisure and tourist oriented population of city centers might be more open minded to changing their eating plans, or to develop them upon reception of the coupon. These points warrant further clarification in order to discern whether our findings can be generalized to other cities alike. Taken together, these findings suggest that marketing managers need to seriously consider the suburban area in which they launch couponing campaigns. Seen the results of this study a 'one-is-good-for-all-strategy' is not advisable in couponing campaigns.

## 6. Limitation and Conclusion

A drawback of this article is that it does not give an insight into absolute redemption rates that can be expected from mobile advertisement channels. Redemption typically varies substantially by distribution method and also by product category (Reibenstein;Traver, 1982). In the experiment reported we used manual distribution of coupons in the street instead of mobile devices delivering ad messages. We also only tested our hypotheses on the basis of one brand and one product category. We therefore could not control for the diverse potential effects of the electronic channel, such as customers not recognizing an incoming message or even negative arousal in response to messages perceived as spam. We also gained no comparative insights into the role of proximity for other product categories than fast food. We also could have inserted more individual propensities to redeem coupons, such as "coupon proneness" in contrast to the value consciousness that we investigated here (Lichtenstein et al., 1990). However, the benefit of our approach resides in the high controllability of the context variables independent of technology pitfalls. The relationships and dynamics observed can serve as a valuable insight for the design of mobile marketing campaigns and couponing campaigns in general and as a starting point to better understand context dynamics in mobile commerce.

Yet, summing up, this article is also the first to our knowledge that systematically investigates the impact of proximity on couponing campaigns and the interplay between proximity and other context variables, such as financial compensation, and neighborhood. We find that proximity of coupon delivery to the place of redemption has a significant effect on campaign success. We observe that proximity can compensate for financial incentives offered to consumers, but also note the effect of powerfully combining proximity with financial gain. With regard to all variables investigated we find that neighborhood plays a crucial role for the strength of effects. All in all, we were able to demonstrate the importance of situation and context for the success of coupon-based marketing. This also makes a strong case for adaptive mobile marketing campaigns and shows which variables to respect when running them.

## 7. References

Agre, P. E. (2001), "Changing Places: Contexts of Awareness in Computing", *Human-Computer Interaction*, Vol. 16 No. 2, pp. 177-192.

Association, N. R. 2009. "2009 Restaurant Industry Pocket Factbook." National Restaurant Association.

Balasubramanian, S., R. A. Peterson and S. L. Jarvenpaa (2002), "Exploring the implications of m-commerce for markets and marketing", *Journal of the Academy of Marketing Science,* Vol. 30 No. 4, pp. 348-362.

Bradley, N. A. and M. D. Dunlop (2005), "Toward a Multidisciplinary Model of Context to Support Context-Aware Computing", *Human-Computer Interaction,* Vol. 20. pp. 403-446.

Brumbaugh, A. M. and J. A. Rosa (2009), "Perceived Discrimination, Cashier Metaperceptions, Embarrassment, and Confidence as Influencers of Coupon Use: An Ethnoracial–Socioeconomic Analysis", *Journal of Retailing,* Vol. 85 No. 3, pp. 347–362.

Chakraborty, G. and C. Cole (1991), "Coupon Characteristics and Brand Choice", *Psychology and Marketing,* Vol. 8 No. 3, pp. 145-159.

Chen, S.-F., S., B. Kent and L. MonroeYung-Chien (1998), "The effects of framing price promotion messages con consumers' perceptions and purchase intentions", *Journal of Retailing,* Vol. 74 No. 3, pp. 353-372.

CouponInfoNow.com. (2010), "Coupon Use Rises 27%, Drives Redemption to 3.3 Billion in 2009." Immar.Inc.

Coutaz, J., J. L. Crowley, S. Dobson and D. Garlan (2005), "Context is Key", *Communications of the ACM,* Vol. 48 No. 3, pp. 49-53.

Submitted draft, April 2010

Dickinger, A. and M. Kleijnen. 2009. "Situational Factors as Drivers of Mobile Services." in Proceedings of the 38th Conference of the European Marketing Academy (EMAC). Nantes.

Ferris, M. (2007), "Insights on Mobile Advertising, Promotion, and Research", *Journal of Advertising Research,* Vol. 7 No. 1, pp. 28-37.

Ferscha, A., W. Swoboda and C. Wimberger. 2009. "En passant Coupon Collection." in Proceedings of the 2nd International Workshop on Pervasive Advertising. Lübeck, Germany.

Ha, H. H., J. S. Hyun and J. H. Pae (2006), "Consumers' "mental accounting" in response to unexpected price savings at the point of sale", *Marketing Intelligence and Planning,* Vol. 24 No. 4, pp. 406 - 416.

Inman, J. and L. McAlister (1994), "Do Coupon Expiration Dates Affect Consumer Behavior?", *Journal of Marketing Research,* Vol. 31 No. 3, pp. 423-428.

Leek, S. and G. Christodoulides (2009), "Next-Generation Mobile Marketing: How Young Consumers React to Bluetooth-Enabled Advertising", *Journal of Advertising Research,* Vol. 49 No. 1, pp. 44-53.

Leppäniemi, M. and H. Karjaluoto (2005), "Factors influencing consumers' willingness to accept mobile advertising: a conceptual model", *International Journal of Mobile Communications,* Vol. 3 No. 3, pp. 197-213.

Lichtenstein, D. R., R. G. Netemeyer and S. Burton (1990), "Distinguishing Coupon Proneness From Value Consciousness: An Acquisition-Transaction Utility Theory Perspective", *Journal of Marketing,* Vol. 54 No. 3, pp. 54-67.

Marketing Week. 2001. "Mobile ad industry survey." in Marketing Week.

Menard, S. (1995), *Applied Logistic Regression Analysis*, Sage Publications. Thousand Oaks,CA.

Mobile Marketing Association. 2007. "Understanding Mobile Marketing - Technology & Reach." ed. Mobile Marketing Association. Denver.

Musalem, A., E. T. Bradlow and J. S. Raju (2008), "Who's Got the Coupon? Estimating Consumer Preferences and Coupon Usage from Aggregate Information", *Journal of Marketing Research,* Vol. 45 No. 6, pp. 715-730.

O'Donoghue, T. and M. Rabin (2000), "The economics of immediate gratification", *Journal of Behavioral Decision Making,* Vol. 13 No. 2, pp. 233 - 250.

Reibenstein, D. J. and P. A. Traver (1982), "Factors Affecting Coupon Redemption Rates", *Journal of Marketing,* Vol. 46 No. 4, pp. 102-113.

Schlosser, E. (2001), *Fast Food Nation: The Dark Side of the All-American Meal*, Harper Perennial. New York.

Thaler, R. (1983), "Transaction Utility Theory", *Advances in Consumer Research,* Vol. 10 No. pp. 296-301.

The Economist. 2007. "Location, location, location." In The Economist. London.

Ward, R. W. and J. E. Davis (1978), "Coupon Redemption", *Journal of Advertising Research,* Vol. 18 No. pp. 51-58.